\documentstyle[12pt]{article}

\newcommand{\dst}{\displaystyle}

\def\a{\alpha }

\begin{document}

\begin{center}
{\Large\bf  "Fixed Point" QCD Analysis of the
CCFR Data on Deep Inelastic Neutrino-Nucleon Scattering}\\[0.2cm]
\end{center}
\vskip 2cm

\begin{center}
{\bf Aleksander V. Sidorov
}
\\
{\it Bogoliubov Theoretical Laboratory\\
 Joint Institute for Nuclear Research\\
141980 Dubna, Russia\\
 E-mail: sidorov@thsun1.jinr.dubna.su}
\vskip 0.5cm

{\bf Dimiter B. Stamenov   \\
{\it Institute for Nuclear Research and Nuclear Energy \\
Bulgarian Academy of Sciences\\
Boul. Tsarigradsko chaussee 72, Sofia 1784, Bulgaria\\
 E-mail:stamenov@bgearn.bitnet}}\\
\end{center}

\vskip 0.3cm
\begin{abstract}
The results of LO {\it Fixed point} QCD (FP-QCD) analysis of the CCFR data for
the nucleon structure function $~xF_3(x,Q^2)~$ are presented. The predictions
of FP-QCD, in which $~\alpha_{s}(Q^2)~$ tends to a nonzero coupling constant
$~\alpha_{0}~$ as $~Q^2\to \infty~$ ,\\ are in good agreement with the data.
The description of the data is even better than that in the case of LO QCD. The
FP-QCD parameter $~\alpha_{0}~$ is determined with a good accuracy:
$~\alpha_{0} = 0.198\pm 0.009~$. Having in mind the recent QCD fits to the same
data we conclude that unlike the high precision and large $~(x,Q^2)~$
kinematic range of the CCFR data they cannot discriminate between QCD and
FP-QCD
predictions for $~xF_3(x,Q^2)~$.\\
\end{abstract}
\vskip 0.5 cm
\newpage

{\bf 1. Introduction.}
\vskip 4mm
The progress of perturbative Quantum Chromodynamics (QCD) in the description
of the high energy physics of strong interactions is considerable. The QCD
predictions are in good quantitative agreement with a great number of data on
lepton-hadron and hadron-hadron processes in a large kinematic region (e.g. see
reviews \cite{Altarelli} and references therein).
Despite of this success of QCD,
we consider
that it is useful and reasonable to put the question: Do the present data fully
exclude the so-called {\it fixed point}  (FP) theory models \cite{Pol} ? \\

We remind that these models are not asymptotically free. The effective coupling
constant $~\alpha_{s}(Q^{2})~$ approaches for $~Q^{2}\to \infty~$ a
constant value $~\alpha_{0}\ne 0~$
(the so-called fixed point at which the Callan-
Symanzik $\beta$-function $~\beta(\alpha_{0}) = 0~$). Using the assumption
that $~\alpha_{0}~$ is small one can make predictions for the physical
quantities in the high energy region, as well as in QCD, and confront them to
 the
experimental data. Such a test of FP theory models has been
made  \cite{GR,BS}
by using the data of deep inelastic lepton-nucleon experiments started by
the SLAC-MIT group \cite{SLAC} at the end of the sixties and performed in
seventies \cite{data70}. It was shown that\\

$~i$) the predictions of the FP theory models with {\it scalar} and {\it non-
colored (Abelian) vector} gluons {\it do not agree} with the data\\
\vskip -5mm
ii) the data {\it cannot distinguish} between different forms of scaling
violation predicted by QCD and the so-called {\it Fixed point} QCD (FP-QCD), a
theory with {\it colored vector} gluons, in which the effective coupling
constant $~\alpha_{s}(Q^{2})~$ does
not vanish when $Q^{2}$ tends to infinity.\\

We think there are two reasons
to discuss again the predictions of FP-QCD. First
of all, there is evidence from the non-perturbative lattice
calculations \cite{Pat} that the $\beta$- function in QCD vanishes
at a nonzero coupling
$~\alpha_{0}~$ that is small. (We remind that the structure of the
$\beta$-function can be studied only by non-perturbative methods.) Secondly,
in the last years the accuracy and the kinematic region of deep inelastic
scattering data became large enough, which makes us hope that discrimination
between QCD and FP-QCD could be performed.\\

In this paper, we present a leading order {\it Fixed point} QCD analysis of the
CCFR data \cite{prep1}. They are most precise data for
the structure function
$~xF_3(x,Q^2)~$. This structure function is pure non-singlet and the results of
analysis are independent of the assumption on the shape of gluons. To
analyze the data the method \cite{Kriv} of reconstruction
of the structure functions
from their Mellin moments is used. This method is based on the Jacobi -
polynomial expansion \cite{Jacobi} of the structure functions.
 In \cite{KaSi} this method has
been already applied to the QCD analysis of the CCFR data.\\

{\bf 2. Method and Results of Analysis.}
\vskip 4mm
Let us start with the basic formulas needed for our analysis.\\

The Mellin moments of the structure function $~xF_3(x,Q^2)~$ are defined as:
\begin{equation}
M_n^{NS}(Q^2)=\int_{0}^{1}dxx^{n-2}xF_{3}(x,Q^2)~,
\label{mom}
\end{equation}
where $~n=2,3,4,...~$.

In FP-QCD the $~Q^2~$ evolution of the non-singlet moments at large  $~Q^2~$ is
given by
\begin{equation}
M_{n}^{NS}(Q^2)
 =M_{n}^{NS}(Q_{0}^2) \left [ \frac{Q_0^{2}}
 {Q^{2}}    \right ]^{\frac{1}{2}\gamma^{NS}_{n}(\alpha_0)}~,
\label{mfp}
\end{equation}
where the anomalous dimensions $~\gamma^{NS}_{n}~$ are determined by its fixed
point value
\begin{equation}
\gamma^{NS}_{n}(\alpha_{0}) = {\alpha_{0} \over {4\pi}}\gamma^{(0)NS}_{n} +
({\alpha_{0} \over {4\pi}})^2\gamma^{(1)NS}_{n} + ... ,
\label{g0}
\end{equation}
and
\begin{equation}
\gamma^{(0)NS}_{n} ={8\over 3}[1 - {2\over n(n+1)} + 4\sum_{j=2}^{n}
{1\over j}]~.
\label{goa0}
\end{equation}

The $n$ dependence of $~\gamma^{(0)NS}_{n},~~ \gamma^{(1)NS}_{n}~$, etc. is
exactly the same as in QCD. However, the $~Q^2~$ behaviour of the moments is
different. In contrast to QCD, the Bjorken scaling for the moments of the
structure functions is broken by powers in $~Q^2~$.\\

In the LO approximation of FP-QCD we have for the moments of $~xF_3(x,Q^2)$:

\begin{equation}
M_{n}^{NS}(Q^2)
 =M_{n}^{NS}(Q_{0}^2) \left [ \frac{Q_0^{2}}
 {Q^{2}}    \right ]^{\frac{1}{2}d^{NS}_{n}}~,
\label{mfplo}
\end{equation}
where
\begin{equation}
 d_n^{NS} = \frac{\alpha_0}{4\pi}\gamma^{(0)NS}_n
\label{dn}
\end{equation}
and $~\alpha_{0}~$ is a free parameter, to be determined from experiment.\\

Having in hand the moments (\ref{mfplo}) and following the method
\cite{Kriv,Jacobi}, we can write
 the structure
function $~xF_3~$ in the form:
\begin{equation}
xF_{3}^{N_{max}}(x,Q^2)=x^{\a}(1-x)^{\beta}\sum_{n=0}^{N_{max}}\Theta_n ^{\a ,
 \beta}
(x)\sum_{j=0}^{n}c_{j}^{(n)}{(\a ,\beta )}
M_{j+2}^{NS} \left ( Q^{2}\right ),   \\
\label{e7}
\end{equation}
where $~\Theta^{\alpha \beta}_{n}(x)~$ is a set of Jacobi polynomials and
$~c^{n}_{j}(\alpha,\beta)~$ are coefficients of the series of
$~\Theta^{\alpha,\beta}_{n}(x)~$ in powers in x:
\begin{equation}
\Theta_{n} ^{\a , \beta}(x)=
\sum_{j=0}^{n}c_{j}^{(n)}{(\a ,\beta )}x^j .
\label{e9}
\end{equation}

$N_{max},~ \alpha~$ and $~\beta~$ have to be chosen so as to
achieve the fastest convergence of the series in the R.H.S.
of Eq.(\ref{e7}) and to
reconstruct $~xF_3~$ with
the accuracy required. Following the results of \cite{Kriv} we use $~\alpha =
 0.12~,
{}~\beta = 2.0~$ and $~N_{max} = 12~$. These numbers guarantee accuracy better
than $~10^{-3}~$.\\

Finally we have to parametrize the structure function $~xF_3~$ at some fixed
value
of $~Q^2 = Q^2_{0}~$. Following \cite{KaSi}, where analysis of the same data is
 done
in the framework of QCD, we choose $~xF_3(x,Q^2)~$ in the simplest form:
\begin{equation}
  xF_{3}(x,Q_0^2)=Ax^{B}(1-x)^{C}~.
\label{e10}
\end{equation}

The parameters A, B and C in Eq. (\ref{e10}) and the FP-QCD parameter
$~\alpha_{0}~$
are free parameters which are determined by the fit to the data.\\

To avoid the influence of higher--twist effects and the target mass
corrections, we have used only the experimental points in the plane $~(x,Q^2)~$
with $~10 < Q^2\leq 501~(GeV/c)^2~$. This cut corresponds to the following
$~x~$ range:$~0.015\leq x \leq 0.65~$.\\

The results of the fit are presented in Table 1. In all fits only statistical
errors are taken into account.
It is seen from the Table that the values of $~\alpha_{0}~$ and
$~\chi^2_{d.f.}~$ are not sensitive to the particular choice of $~Q^2_{0}~$.
This is an indication of the stability and the self-consistence of the method
used.\\

The values of $~\chi^2_{d.f.}~$ presented in Table 1 are slightly smaller
than those obtained in the LO QCD analysis \cite{KaSi} of
the CCFR data and indicate
a good description of the data. The values of the parameters A, B and C
are in agreement with the results of \cite{KaSi}.\\
\vskip 0.2 cm
\begin{tabular}{|c|c|c|c|c|c|c|} \hline
$Q_0^2 $&$\chi^2_{d.f.}$& $\alpha_0$ & A & B & C & GLS  \\
$(GeV/c)^2$&               &            &   &   &   & sum rule\\  \hline
  3  & 82.2/61& .198$\pm$.009 & 6.50$\pm$.18 & .768$\pm$.013 &3.44$\pm$.04 &
 2.539$\pm$.111\\
 10  & 82.9/61& .198$\pm$.009 & 5.93$\pm$.15 & .722$\pm$.012 &3.56$\pm$.034&
 2.564$\pm$.106\\
 20  & 83.5/61& .198$\pm$.009 & 5.62$\pm$.15 & .696$\pm$.012 &3.64$\pm$.032&
 2.580$\pm$.111\\
 50  & 84.5/61& .198$\pm$.009 & 5.24$\pm$.14 & .663$\pm$.012 &3.73$\pm$.031&
 2.605$\pm$.115\\
100  & 85.3/61& .198$\pm$.009 & 4.96$\pm$.13 & .638$\pm$.012 &3.80$\pm$.029&
 2.626$\pm$.117\\ \hline
\end{tabular}
\vskip 4mm
\begin{tabular}{cl}
{\bf Table 1.}& The results of the LO FP-QCD fit
to the CCFR $~xF_3~$ data for  $f=4$.\\
&$\chi^2_{d.f.}$ is the
$\chi^2$-parameter normalized to the degree of freedom $d.f.$.
\end{tabular}
\vskip 0.4 cm

Previous estimations \cite{BS} of the FP-QCD parameter $~\alpha_{0}~$  based on
 the
analysis of SLAC deep inelastic electron-proton data provide a large region for
possible values of $~\alpha_{0}~$:\\
\begin{equation}
       0.1  < \alpha_0 < 0.4~.
\label{aBS}
\end{equation}

Now $~\alpha_{0}~$ is determined from the CCFR data with a good accuracy in the
above interval:
\begin{equation}
\alpha_{0} = 0.198\pm0.009~.
\label{a0res}
\end{equation}

The value of the Gross-Llewellyn Smith (GLS) sum rule has been calculated at
different values of $~Q^2_{0}~$ as the first moment of $~xF_3(x,Q^2_{0})$
\begin{equation}
 \dst{
GLS(Q^2_0)=\int_{0}^{1}\frac{dx}{x}A(Q^2_0)     x^{{B}(Q^2_0)}
(1-x)^{{C}(Q^2_0)}        }
\label{GLS}
\end{equation}
with an accuracy about 4\%. These values (see Table 1) are in good
agreement with LO QCD results of \cite{KaSi}.\\

{\bf 3. Summary.}
\vskip 4mm
The CCFR deep inelastic nucleon scattering data have been analyzed in the
framework of the {\it Fixed point} QCD. It was demonstrated that the data for
the nucleon structure function $~xF_3(x,Q^2)~$ are in good agreement with
the LO predictions of this theory model using the assumption that the
{\it fixed point} coupling $~\alpha_{0}~$ is small. In contrast to the results
of the fits to the previous generations of deep inelastic lepton-nucleon
experiments, the value of this constant was determined with a good accuracy:
$~\alpha_{0} = 0.198\pm 0.009~$. This value of $~\alpha_{0}~$ is consistent
with the assumption that $~\alpha_{0}~$ is small.\\

In conclusion, we find that the CCFR data, the most precise data on deep
inelastic scattering at present, {\it do not eliminate} the FP-QCD and
therefore other tests have to be made in order to distinguish between QCD and
FP-QCD.\\
\vskip 4mm
{{ \bf Acknowledgement}}
\vskip 3mm
We are grateful to  M. H. Shaevitz for providing us with
the CCFR data. One of us (D.S.) would like to thank also the Bogoliubov
Theoretical Laboratory for hospitality at the JINR in Dubna where this work
was completed.\\

This research was partly supported by INTAS (International
Association for the Promotion of Cooperation with Scientists from the
Independent States of the Former Soviet Union) under Contract nb 93-1180, by
the Russian Fond for Fundamental Research Grant N 94-02-04548-a
and by Bulgarian Science Foundation under Contract \mbox{F 16.}\\

\end{document}